\PassOptionsToPackage{dvipsnames}{xcolor}
\PassOptionsToPackage{hyphens}{url}

\documentclass[sigconf]{acmart}
\usepackage{balance}
\usepackage{tikz}
\usepackage[relative,overlay]{textpos}
\usepackage{fancyvrb}
\usepackage{listings}
\usepackage{breakurl}

\newcommand*\circled[1]{\tikz[baseline=(char.base)]{
    \node[shape=circle,fill=black, text=white,draw,inner sep=2pt] (char) {#1};}}

\lstdefinestyle{base}{
  language=Python,
  emptylines=1,
  breaklines=true,
  basicstyle=\ttfamily\color{black},
  moredelim=**[is][\color{red}]{@}{@},
  moredelim=**[is][\color{blue}]{|}{|}
}

\def\checkmark{\tikz\fill[scale=0.4](0,.35) -- (.25,0) -- (1,.7) -- (.25,.15) -- cycle;} 

\AtBeginDocument{%
  \providecommand\BibTeX{{%
    \normalfont B\kern-0.5em{\scshape i\kern-0.25em b}\kern-0.8em\TeX}}}

\copyrightyear{2023}
\acmYear{2023}
\setcopyright{rightsretained}
\acmConference[ICER '23 V1]{Proceedings of the 2023 ACM Conference on International Computing Education Research V.1}{August 7--11, 2023}{Chicago, IL, USA}
\acmBooktitle{Proceedings of the 2023 ACM Conference on International Computing Education Research V.1 (ICER '23 V1), August 7--11, 2023, Chicago, IL, USA}
\acmDOI{10.1145/3568813.3600142}
\acmISBN{978-1-4503-9976-0/23/08}

\begin{document}

\title[Large Language Models (GPT-4) No Longer Struggle to Pass Assessments]{Thrilled by Your Progress!\\ Large Language Models (GPT-4) No Longer Struggle to Pass Assessments in Higher Education Programming Courses}

\author{Jaromir Savelka}
\orcid{0000-0002-3674-5456}
\affiliation{%
  \institution{Carnegie Mellon University}
  \city{Pittsburgh}
  \state{PA}
  \country{USA}
}
\email{jsavelka@cs.cmu.edu}

\author{Arav Agarwal}
\affiliation{%
  \institution{Carnegie Mellon University}
  \city{Pittsburgh}
  \state{PA}
  \country{USA}
}
\email{arava@andrew.cmu.edu}

\author{Marshall An}
\affiliation{%
  \institution{Carnegie Mellon University}
  \city{Pittsburgh}
  \state{PA}
  \country{USA}
}
\email{haokanga@andrew.cmu.edu}

\author{Chris Bogart}
\affiliation{%
  \institution{Carnegie Mellon University}
  \city{Pittsburgh}
  \state{PA}
  \country{USA}
}
\email{cbogart@andrew.cmu.edu}

\author{Majd Sakr}
\affiliation{%
  \institution{Carnegie Mellon University}
  \city{Pittsburgh}
  \state{PA}
  \country{USA}
}
\email{msakr@cs.cmu.edu}

\renewcommand{\shortauthors}{Savelka et al.}

\begin{abstract}
  This paper studies recent developments in large language models' (LLM) abilities to pass assessments in introductory and intermediate Python programming courses at the postsecondary level. The emergence of ChatGPT resulted in heated debates of its potential uses (e.g., exercise generation, code explanation) as well as misuses in programming classes (e.g., cheating). Recent studies show that while the technology performs surprisingly well on diverse sets of assessment instruments employed in typical programming classes the performance is usually not sufficient to pass the courses. The release of GPT-4 largely emphasized notable improvements in the capabilities related to handling assessments originally designed for human test-takers. This study is the necessary analysis in the context of this ongoing transition towards mature generative AI systems. Specifically, we  report the performance of GPT-4, comparing it to the previous generations of GPT models, on three Python courses with assessments ranging from simple multiple-choice questions (no code involved) to complex programming projects with code bases distributed into multiple files (599 exercises overall). Additionally, we analyze the assessments that were not handled well by GPT-4 to understand the current limitations of the model, as well as its capabilities to leverage feedback provided by an auto-grader. We found that the GPT models evolved from completely failing the typical programming class' assessments (the original GPT-3) to confidently passing the courses with no human involvement (GPT-4). While we identified certain limitations in GPT-4's handling of MCQs and coding exercises, the rate of improvement across the recent generations of GPT models strongly suggests their potential to handle almost any type of assessment widely used in higher education programming courses. These findings could be leveraged by educators and institutions to adapt the design of  programming assessments as well as to fuel the necessary discussions into how programming classes should be updated to reflect the recent technological developments. This study provides evidence that programming instructors need to prepare for a world in which there is an easy-to-use widely accessible technology that can be utilized by learners to collect passing scores, with no effort whatsoever, on what today counts as viable programming knowledge and skills assessments.
\end{abstract}

\ccsdesc[500]{Social and professional topics~Computing education}
\ccsdesc[400]{Social and professional topics~Student assessment}
\ccsdesc[500]{Computing methodologies~Artificial intelligence}
\ccsdesc[400]{Computing methodologies~Natural language processing}

\keywords{AI code generation, introductory and intermediate programming, Multiple-choice question answering, MCQ, coding exercises, generative pre-trained transformers, GPT, Python course, programming knowledge assessment, ChatGPT, Codex, GitHub Copilot, AlphaCode}


\maketitle

\section{Introduction}
Rapidly increasing capabilities of large language models (LLM) keep challenging established practices in various contexts, including computer science and information technology (CS/IT) education. There are important unanswered questions related to (i) curricular changes needed to accommodate the new reality, (ii) excessive learners' reliance on LLMs in engaging with learning materials, assignments, and assessments, as well as (iii) considerable uncertainty as to how the future of CS/IT profession(al)s look like. Hence, the all-important concern shared by many CS/IT educators as to what are the skills and knowledge the learners in CS/IT programs need in order to have successful and meaningful careers. Perhaps, a more immediate question that likely occupies minds of many instructors is how to assess learners' skills and knowledge in the presence of ubiquitous tools (e.g., ChatGPT,\footnote{OpenAI: ChatGPT. Available at: \url{https://chat.openai.com/} [Accessed 2023-03-20]} GitHub Copilot\footnote{GitHub Copilot: Your AI pair programmer. Available at: \url{https://github.com/features/copilot} [Accessed 2023-03-20]}) that could be easily utilized to pass the assessments (at least partially).

While it is difficult to provide definitive or even satisfactory answers to questions posed above it is of utmost importance to build and maintain a body of empirically validated knowledge that would facilitate deep and meaningful discussions on these topics. Indeed, there has been a growing body of scholarship focused on understanding the capabilities of LLMs, as well as their limitations, in the context of programming education (see Section \ref{sec:related_work}). The recent release of GPT-4 poses a challenge for the existing work that needs to be confirmed or updated to account for this seemingly more powerful technology. The issue is especially pressing when it comes to what we know about the capabilities of GPT models to handle assessments that were originally designed for a human test-taker. This is because GPT-4 appears to perform much better on academic and professional exams when compared to the preceding GPT-3.5 generation. The technical report \cite{openai2023gpt4} made available with the GPT-4 release lists 34 such exams, including various Graduate Record Examination (GRE) tests,\footnote{ETS: The GRE General Test. Available at: \url{https://www.ets.org/gre/test-takers/general-test/about.html} [Accessed 2023-03-20]} SAT Math,\footnote{SAT Suite of Assessments. Available at: \url{https://satsuite.collegeboard.org/sat/whats-on-the-test/math} [Accessed 2023-03-20]} or a Uniform Bar Exam.\footnote{NCBE: Uniform Bar Examination. Available at: \url{https://www.ncbex.org/exams/ube/} [Accessed 2023-03-20]} Several of the listed exams involve programming tasks (e.g., Leetcode,\footnote{LeetCode. Available at: \url{https://leetcode.com/} [Accessed 2023-03-22]} Codeforces Rating\footnote{Codeforces. Available at: \url{https://codeforces.com/contests} [Accessed 2023-03-22]}) and those too show notably improved performance. Hence, it appears that the current knowledge of the capabilities of the GPT models to handle assessments in programming courses might be outdated.

This paper analyzes the capabilities of the newest state of the art generative pre-trained transformer (GPT-4) to pass typical assessments, i.e., multiple-choice question (MCQ) tests and coding exercises, in introductory and intermediate programming courses. The aim of this paper is to quickly react to the recent release of GPT-4 and assess if and to what extent do the findings presented by similar past studies performed with GPT-3 and 3.5 models still stand. Hence, the focus is not only on the performance of GPT-4 but also on the comparison of its performance to that of the earlier GPT models. To that end we employ a data set comprising of 599 assessments from three currently running Python courses. We asses the outputs of the GPT models as if they were coming from a human learner. This means that we also expose the models to a feedback generated by an auto-grader and provide them with an opportunity to iterate on the solution. This is all done in a manner ensuring that there is no human intervention that could contribute to models successfully passing the assessments. This approach allows us to accurately gauge if the automatically generated solutions would enable a human learner to pass a course provided it would have been their own work. The immediate insight that this study offers is that the danger of learners' over-reliance on GPT models when completing their programming courses' assignments and assessments is a real concern that has to be taken seriously.

To investigate if and how GPT-4 challenges the findings of the prior works related to the capabilities of LLMs to handle diverse types of assessments typically employed in real-world introductory and intermediate Python programming courses at the post-secondary level, we analyzed the following research questions from the prior work in light of the newly released model:

\begin{itemize}
    \item[(RQ1)] To what degree can GPT-4 generate correct answers to MCQs in order to pass an introductory or intermediate course in Python in higher education? \cite{Savelka2023,savelka2023large}
    \item[(RQ2)] Does GPT-4 struggle with programming MCQs containing code snippets that require multi-hop reasoning? \cite{savelka2023large}
    \item[(RQ3)] To what degree can GPT-4 produce solutions to complex coding tasks from instructions in order to pass an introductory or intermediate course in Python in higher education? \cite{Savelka2023,10.1145/3511861.3511863,finnie2023my,denny2023conversing,piccolo2023many}
    \item[(RQ4)] Can GPT-4 successfully utilize feedback to fix solutions of coding tasks? \cite{Savelka2023}
\end{itemize}

By carrying out this work, we provide the following contributions to the CS education research community. To our best knowledge, this is the first comprehensive study that:

\begin{itemize}
    \item[(C1)] Measures performance of the GPT-4 model on diverse assessment instruments from Python programming courses, updating and extending the current body of knowledge that has been developed on experiments with GPT-3 models.
    \item[(C2)] Offers a detailed in-depth analysis of common properties of MCQs and coding tasks that are answered incorrectly by GPT-4.
\end{itemize}

\section{Related Work}
\label{sec:related_work}

\textbf{GPT-4.}
Given the recent arrival of GPT-4, there have been few studies of the implications of the new model in education as of the writing of this paper. OpenAI's technical report states performance of GPT-4 on numerous tasks across diverse domains. Of particular importance are the 92.0\% success rate of GPT-4 on grade-school mathematics questions using 5-shot examples and chain-of-thought prompting, solving 31/41 Leetcode easy and 21/80 Leetcode medium exercises, and significant success across several quantitative AP and competitive mathematics exams \cite{openai2023gpt4}. Katz et al. demonstrate that GPT-4 achieves 297 points on the Uniform Bar Exam (UBE), passing the bar exam and, in the authors words ``by a significant margin'' \cite{katz2023gpt}. Lastly, Jiao et al. consider the performance of GPT-4 on academic translation tasks, demonstrating that the ChatGPT service achieves significantly better performance compared to existing commercial translation products \cite{jiao2023chatgpt}. This paper falls in line with such work, conducting a rigorous evaluation of the recently-released GPT-4 model when applied to typical introductory and intermediate programming assessments. We demonstrate that the gains in performance observed in other domains extend to the programming education as well.

\textbf{GPT Performance on Programming MCQs.}
Savelka et al. evaluated the capability of \verb|text-davinci-003|, to pass a diverse set of assessment instruments, including MCQs, in the realistic context of full-fledged programming courses \cite{Savelka2023}. They found that the then current GPT models were not capable of  passing the full spectrum of assessments typically involved in a Python programming course (below 70\% on even entry-level modules); but a straightforward application of these models could enable a learner to obtain a non-trivial portion of the overall available score~(over 55\%) in introductory and intermediate courses alike. They also observed that an important limitation of the  GPT models was their apparent struggle with activities that required multi-hop reasoning, and that there appeared to be a difference in success rate between MCQs that contained a code snippet and those that did not \cite{Savelka2023,savelka2023large}. In this work, we re-examine those findings in the light of the more powerful model released since their publication. We find that the conclusions about the models not being able to pass the courses no longer hold. However, some of the limitations identified in the prior work still hold.

\textbf{GPT Performance on MCQs in Other Domains.}
Robinson et al. apply InstructGPT \cite{ouyang2022training} and Codex to OpenBookQA~\cite{mihaylov2018can}, StoryCloze \cite{mostafazadeh2016corpus}, and RACE-m \cite{lai2017race} data sets which focus on multi-hop reasoning, recall, and reading comprehension, reporting 77.4-89.2\% accuracy \cite{robinson2022}. In some cases, GPT can generate code when applied to programming assignments in higher education courses. Drori and Verma used Codex to write Python programs to solve 60 computational linear algebra MCQs, reporting 100\% accuracy \cite{https://doi.org/10.48550/arxiv.2111.08171}. Others have used GPT models to solve various MCQ-based exams, including the United States Medical Licensing Examination (USMLE), with accuracy around 50\% \cite{kung2022performance,Gilson2022HowWD,Lievin2022CanLL}, the Multistate Bar Examination (MBE) \cite{bommarito2022gpt,katz2023gpt}, and the American Institute of Certified Public Accountants' (AICPA) Regulation (REG) exam \cite{bommarito2023gpt}. Although, GPT can often answer questions \emph{about} systems and rules, it is especially challenged by tasks that involve \emph{applying} them and reasoning about their implications in novel examples. Hendryks et al. created data set that includes a wide variety of MCQs across STEM, humanities and arts, with GPT-3 performing at levels above 50\% for subjects such as marketing and foreign policy, but below 30\% for topics such as formal logic~\cite{hendrycks2022}. They found that the model performed particularly poorly in quantitative subjects. For example, in Elementary Mathematics they note that GPT can answer questions \emph{about} arithmetic order of operations (e.g., that multiplications are performed before additions), but it cannot correctly answer questions that require \emph{applying} this concept. They also note that GPT performance is not necessarily correlated with how advanced the topic is for humans, doing better at College Mathematics than Elementary Mathematics. Finally, they noted that GPT does poorly on tests of legal and moral reasoning~\cite{hendrycks2022}. Lu et al. studied GPT models' performance on a large data set consisting of 21,208 MCQs on topics in natural science, social science, and language  \cite{pan2022}. They prompted the models to produce an explanation along with its answer and reported  1-3\% improvement in accuracy (74.04\%). In this work, we do not adopt the approach and, hence, leave space for future work as it appears quite promising and definitely applicable in the context of programming MCQs.

\textbf{GPT Performance on Coding Assessments.}
There is a growing body of related work on GPT models' capabilities in solving educational programming tasks by generating code and text. Finnie-Ansley et al. evaluated Codex on 23 programming tasks used as summative assessments in a CS1 \cite{10.1145/3511861.3511863} and CS2 \cite{finnie2023my} programming courses. Denny et al. focused on the effects of prompt engineering when applying Copilot to a set of 166 exercises from the publicly available CodeCheck repository~\cite{denny2023conversing}. Jalil et al. evaluated the performance of ChatGPT on content from five chapters of software testing curricula, reporting a 55.6\% accuracy in their assessment \cite{jalil2023}. Our paper extends the existing body of work, most importantly by using the more powerful GPT-4 model. Piccolo et al. used 184 programming exercises from an introductory bioinformatics course to evaluate the extent to which ChatGPT can successfully complete basic to moderate level programming tasks, reporting the success rate of 75.5\% on the first attempt and 97.3\% when provided with feedback \cite{piccolo2023many}. Several studies focus on collaboration between a human learner and GPT-based assisting tools (e.g., Copilot). Kazemitabaar et al. studied learners using OpenAI's Codex during traditional code creation tasks, and demonstrated the use of Codex did not harm their performance, with experienced students performing significantly better \cite{NoviceCodex23}. Leinonen et al. used Codex to generate more readable error messages for learners to use for project-level debugging, suggesting that model-created explanations can serve as effective scaffolding for students learning to program \cite{leinonen2023pem}. Prather et al. examined how novices interact with these tools, observing that novices struggle to understand and use Copilot \cite{prather2023s}.

\textbf{GPT and Computing Education.}
Besides the work focused on how well LLMs do in various tasks meant to be performed by human learners, there is also a growing body of work on using LLMs to support computing education. Sarsa et al. used \verb|code-davinci-001| to generate 240 introductory programming exercises, along with tests, sample solutions and explanations, reporting that over 75\% of the generated exercises were novel and suitable for use in a university setting \cite{sarsa2022}. Macneil et al. integrated LLM-generated code explanations into an interactive e-book and compared several different explanation types, such as line-by-line or summarization-oriented explanations, reporting students using line-by-line explanations the most despite them being considered the least useful according to the students \cite{macneilebook2023}. Leinonen et al. compared learner-authored code explanations with those generated by GPT-3, showing that students perceived GPT-3 generated explanations as more readable of the two \cite{leinonen2023comparing}.


\textbf{GPT Performance on Coding Tasks in Professional Settings.}
Outside of the educational context, there have been studies exploring GPT's capabilities on competitive and interview programming tasks. Chen et al. released the HumanEval data set where Codex achieved 28.8\% success rate on the first attempt and 72.3\% when allowed 100 attempts \cite{https://doi.org/10.48550/arxiv.2107.03374}. Li et al. report Deepmind's AlphaCode performance on Codeforces competitions,\footnote{Codeforces. Available at: \url{https://codeforces.com/contests} [Accessed 2023-01-22]} achieving a 54.3\% ranking amongst 5,000 participants \cite{doi:10.1126/science.abq1158}. Karmakar et al. reported 96\% pass rate for Codex on a data set of 115 programming problems from HackerRank\footnote{HackerRank. Available at: \url{https://www.hackerrank.com/} [Accessed 2023-01-22]} \cite{Karmakar2022CodexHH}. Nguyen and Nadi reported Copilot's effectiveness on LeetCode\footnote{LeetCode. Available at: \url{https://leetcode.com/} [Accessed 2023-01-22]} problems, achieving 42\% accuracy \cite{9796235}. Perry et al. explored the security implications of using Copilot \cite{perry2022users}.

Program code does more than control computer execution; it also, some argue primarily, serves as communication among developers~\cite{Knuth1984}. Since GPT is a text prediction model trained on code in the context of human discussions about it, the model's representation of code is likely to capture code's \emph{design intent} more strongly than code's \emph{formal properties}. For example, work from multiple studies suggest that models that interpret code depend heavily on function names and input variables \cite{Mohammadkhani2022ExplainableAF,robustness}. Although, models like GPT are not trained to simulate code execution, they can in many cases generate code based on natural language description of the code's intent. Researchers have reported varying success at generating code in response to programming assignments, ranging from Codex's 100\% success generating Python computational linear algebra programs \cite{https://doi.org/10.48550/arxiv.2111.08171}, to 78.3\% on some CS1 programming problems ~\cite{10.1145/3511861.3511863}, to 79\% on the CodeCheck\footnote{CodeCheck: Python Exercises. Available at: \url{https://horstmann.com/codecheck/python-questions.html} [Accessed 2022-01-22]} repository of Python programming problems~\cite{denny2023conversing}.

\textbf{Prompt Engineering.}
It has been well established that LLMs are few-shot learners, capable of answering questions without additional fine-tuning in a zero-shot fashion \cite{brown2020language}. In general, finding the best prompt for a specific task is challenging, with prompts that are semantically similar sometimes providing large differences in performance \cite{zhao2021calibrate}. Despite this difficulty, there have been several advancements in developing techniques for prompt engineering to improve the performance of LLMs. Numerous studies have explored prompts which include a number of examples to demonstrate what the desired output should be \cite{gao2021making,brown2020language,Zhang2023PromptingLL}. However, the current research literature remains inconclusive as to the efficacy of adding examples to natural language prompts, with multiple studies suggesting that the order and number of examples can dramatically influence the performance of LLMs across various tasks \cite{Zhang2023PromptingLL,reynolds2021prompt}.

In introductory CS context, there has been an inquiry into explainable prompt-engineering practices. Denny et al. explored prompting Copilot for CS1 exercises, demonstrating that while prompt engineering can significantly improve the performance of Copilot on CS1 problems, verbose prompts can lead to decreases in model performance \cite{denny2023conversing}. Similar study has been performed on CS2 coding tasks \cite{finnie2023my}.

More recently, there has been significant interest in chain of thought prompting, a technique where an LLM is asked to provide both the answer and the reasoning that lead to the answer in question. This has lead to significant performance gains in symbolic and quantitative reasoning tasks, by forcing the LLM to emulate human reasoning in addition to the answer itself \cite{Wei2022ChainOT}. Recently, researchers have also explored the so called ``least-to-most'' prompting, where a task is decomposed into several sub-problems, which are then answered all at once by the model \cite{Zhou2022LeasttoMostPE}.

\section{Data}
For the purpose of this study, we obtained the data set that was originally used in \cite{Savelka2023,savelka2023large}. The researchers collected assessment exercises from three real-world currently running Python programming courses.

\emph{Python Essentials - Part 1 (Basics)}\footnote{OpenEDG: Python Essentials - Part 1 (Basics). Available at: \url{https://edube.org/study/pe1} [Accessed 2023-03-20]} (\textbf{PE1}) transitions learners from a state of complete programming illiteracy to a level of programming knowledge which allows them to design, write, debug, and run Python programs. There are four units in the course, and one completion (summary) test. The units include:

\begin{enumerate}
    \item Introduction to Python and computer programming,
    \item Data types, variables, basic input-output operations and basic operators
    \item Boolean values, conditional loops, lists, logical and bitwise operators
    \item Functions, tuples, dictionaries and data processing.
\end{enumerate} 

\noindent PE1 employs MCQ assessments. Formative assessments are called quizzes and summative assessments are called tests. Qualitatively, the test MCQs appear to be considerably more challenging than quiz MCQs. The MCQs often include small snippets of code and ask learners to reason about them.

There are 149 questions in PE1. An MCQ may involve a snippet of Python code (\emph{with code}) or it may be expressed fully in natural language (\emph{no code}). For an MCQ, to be considered as \emph{with code} there either is at least one line fully dedicated to computer code, and/or the choices are computer code expressions. Inline mentions of names of functions or variables were not considered as sufficient for an MCQ to be considered \emph{with code}. Out of the 149 MCQs in PE1, 96 have code snippets. The MCQs are further distinguished into the following categories:

\begin{itemize}
     \item \emph{True/False} -- The learner is asked to assess the truthfulness of a single statement.
     \item \emph{Identify True/False Statement} -- The learner is asked to pick one or more choices as either true or false.
     \item \emph{Finish Statement.} -- The learner is asked to complete a statement.
     \item \emph{Output} -- The learner is asked to identify the choice that corresponds to the output of a given snippet of code.
     \item \emph{Fill-in Blanks} -- The learner is asked to fill in a code snippet by selecting the appropriate choice as an answer.
     \item \emph{Other} -- Any MCQ that does not fall into any of the above categories.
\end{itemize}

\noindent Table~\ref{tab:dataset} provides additional details, including the categorization of questions according to their type. Example questions for all the types are shown in Appendix \ref{app:mcqs}.

\begin{table*}
  \caption{MCQ Data Set. Each row provides information about the MCQ assessments each of the courses employ. Each column reports on the distribution of the MCQ types across the courses.}
  \label{tab:dataset}
  \begin{tabular}{l|r|rrrr|rrrrrr|r}
  \toprule
    Course        & Units    & \multicolumn{4}{c|}{MCQ (no code)} & \multicolumn{6}{c|}{MCQ (with code)} & Course\\
                  & (topics) & T/F & Id. T/F & Fin. S. & Other   & T/F & Id. T/F & Fin. S. & Out & Fill-in & Other& Overall\\
  \hline
    PE1           & 4        & 0   & 5       & 23      & 18      & 0   & 5       & 6       & 51  & 0       & 41   &\bf 149 \\
    PE2           & 4        & 0   & 7       & 31      & 10      & 0   & 0       & 21      & 27  & 0       & 52   &\bf 148 \\
    PPP           & 8        & 25  & 32      & 2       & 19      & 23  & 21      & 0       & 32  & 13      & 66   &\bf 233        \\
  \hline
    Type Overall  & 16 &\bf 25 &\bf 44 &\bf 56 &\bf 47 &\bf 23 &\bf 26 &\bf 27 &\bf 110 &\bf 13 &\bf 159 &\bf 530  \\
  \bottomrule
  \end{tabular}
\end{table*}

\emph{Python Essentials - Part 2 (Intermediate)} (\textbf{PE2})\footnote{OpenEDG: Python Essentials - Part 2 (Intermediate). Available at: \url{https://edube.org/study/pe2} [Accessed 2023-01-15]} covers advanced aspects of Python programming, such as modules, packages, exceptions, file processing, or object-oriented programming. Similarly to PE1, the course is organized into four units and it is also equipped with a completion (summary) test. The course units are:

\begin{enumerate}
    \item Modules, packages, and PIP
    \item Strings, String and List Methods, Exceptions
    \item Object-Oriented Programming
    \item Miscellaneous
\end{enumerate}

\noindent Just like PE1, PE2 also employs MCQ assessments exclusively (quizzes and tests). There are 148 questions in PE2 out of which 83 have code snippets. Table~\ref{tab:dataset} has additional details.

Finally, \emph{Practical Programming with Python}\footnote{Sail(): Social and Interactive Learning Platform. Available at: \url{https://sailplatform.org/courses}. [Accessed 2023-03-20]} (\textbf{PPP}) centers around hands-on projects focused on fundamental Python constructs and exposure to software development tools, practices, and real-world applications. The course consists of eight units which include:

\begin{enumerate}
    \item Python basics and introduction to functions
    \item Control flow, strings, input and output
    \item Python data structures,
    \item Object-oriented programming
    \item Software development
    \item Data manipulation
    \item Web scraping and office document processing
    \item Data analysis
\end{enumerate}

\noindent PPP also uses MCQs extensively. However, their influence on learners' passing the course is limited compared to PE1 and PE2. In PPP MCQs are used as inline gating activities meant as formative assessments and graded tests as summative assessments. The contribution of the tests to the overall grade would vary across the PPP offerings but it would rarely exceed 20\%. There are 233 MCQs in PPP (144 with code snippets). Table~\ref{tab:dataset} has additional details about MCQs in PPP.

\begin{table}
  \caption{Coding Activities Data Set. Each row provides information about each project's focus area(s) and the number of tasks and activities they contain. Fund. stands for Python language fundamentals; SW Dev. stands for software development practices (e.g., Test-driven development); Data stands for data processing and analysis (e.g., file formats, databases).}
  \label{tab:projects}
  \setlength{\tabcolsep}{3pt}
  \begin{tabular}{lcccrr}
    \toprule
    Project Topic                                 & Fund.    & SW Dev.  & Data     & Tasks & Acts.\\
    \midrule
    Types, variables, functions                   &\checkmark&\checkmark&          & 4     & 6\\
    Iteration, conditionals,                      &\checkmark&          &          & 4     & 6\\
    \ \ \ strings, basic I/O                            \\
    Lists, sets, tuples and                       &\checkmark&          &\checkmark& 2     & 11\\
    \ \ \ dictionaries                                  \\
    Classes, objects, attributes                  &\checkmark&\checkmark&          & 6     & 8\\
    \ \ \ and methods                                  \\  
    Debugging, refactoring,                       &          &\checkmark&          & 6     & 15\\
    \ \ \ testing and packaging                         \\
    Files and datastores                          &          &          &\checkmark& 3     & 7\\
    Web scraping and office                       &          &          &\checkmark& 4     & 7\\
    \ \ \ document processing                           \\
    Data analysis                                 &          &          &\checkmark& 3     & 9\\
    \bottomrule
                                                  &          &          &          &\bf 32 &\bf 69\\ 
    \cline{5-6}
  \end{tabular}
\end{table}

In comparison to PE1 and PE2, PPP mostly employs the project-based education model \cite{kokotsaki2016project}. Learners individually work on larger programming projects that are subdivided into tasks. The projects require sustained effort that often extends over several days, depending on the proficiency of the learner. All the eight projects are auto-graded. The auto-grader provides learners with feedback that can be utilized for improving the solutions until a project deadline. The score from the projects, tests, and reflections (discussion posts) determines if a learner successfully completes the course. The projects typically contribute around 80\% towards the grade. There are 69 coding activities in PPP (elements of the 32 project tasks). Further details about the projects and their coding activities are reported in Table~\ref{tab:projects}.

\begin{figure}
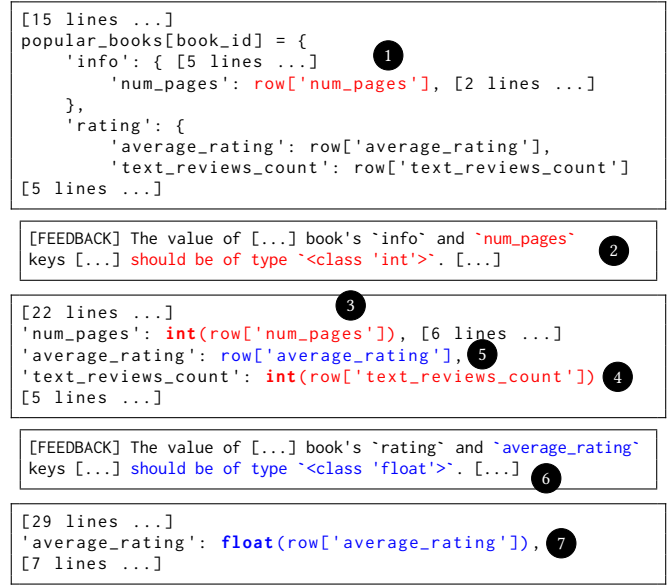

\footnotesize
\begin{lstlisting}[frame=single,style=base]
[15 lines ...]
popular_books[book_id] = {
    'info': { [5 lines ...]
        'num_pages': @row['num_pages']@, [2 lines ...]
    },
    'rating': {
        'average_rating': row['average_rating'],
        'text_reviews_count': row['text_reviews_count']
[5 lines ...]
\end{lstlisting}
\begin{textblock*}{1.4cm}(4.2cm,-2.35cm)
\circled{1}
\end{textblock*}

\begin{Verbatim}[frame=single,commandchars=\\\{\}]
[FEEDBACK] The value of [...] book's `info` and \textcolor{red}{`num_pages`} 
keys [...] \textcolor{red}{should be of type `<class 'int'>`}. [...]
\end{Verbatim}
\begin{textblock*}{1.4cm}(7.2cm,-0.65cm)
\circled{2}
\end{textblock*}

\begin{lstlisting}[frame=single,style=base]
[22 lines ...]
'num_pages': @int(row['num_pages'])@, [6 lines ...]
'average_rating': |row['average_rating']|,
'text_reviews_count': @int(row['text_reviews_count'])@
[5 lines ...]
\end{lstlisting}
\begin{textblock*}{1.4cm}(3.7cm,-1.8cm)
\circled{3}
\end{textblock*}
\begin{textblock*}{1.4cm}(7.25cm,-0.85cm)
\circled{4}
\end{textblock*}
\begin{textblock*}{1.4cm}(5.45cm,-1.15cm)
\circled{5}
\end{textblock*}

\begin{Verbatim}[frame=single,commandchars=\\\{\}]
[FEEDBACK] The value of [...] book's `rating` and \textcolor{blue}{`average_rating`} 
keys [...] \textcolor{blue}{should be of type `<class 'float'>`}. [...]
\end{Verbatim}
\begin{textblock*}{1.4cm}(6.3cm,-0.4cm)
\circled{6}
\end{textblock*}

\begin{lstlisting}[frame=single,style=base]
[29 lines ...]
'average_rating': |float(row['average_rating'])|, 
[7 lines ...]
\end{lstlisting}
\begin{textblock*}{1.4cm}(6.5cm,-0.85cm)
\circled{7}
\end{textblock*}
\caption{An example interaction with auto-grader which recognizes that the num\_pages field (1) should not be of type str and produces corresponding feedback (2). After the flaw is corrected (3), as well as similar one (4) not mentioned in the feedback, another issue with the average\_rating field~(5) is fixed (7) based on the additional feedback (6).}
\label{fig:auto-grader}
\end{figure}

Each project activity is associated with one or more assessments which are high-level rules that need to be met in order for a learner to be awarded with score points. For example, an assessment could require the output JSON file to have specific fields in terms of their names and data types. The auto-grader then uses an extensive battery of detailed tests to ensure the high-level assessment rule is met. The test cases are dynamically generated during the evaluation of each submission. In case one or more tests fail, the learner receives a feedback the aim of which is to (1) clearly explain as to why the assessment rule is not met, and, hence, the score cannot be awarded, and (2) provide a hint on how to iterate on the solution towards a successful outcome. The feedback would typically not provide an extensive enumeration of the failed test cases. Instead, it often focuses on the most prominent one or the first one encountered. The feedback usually does not expose the exact nature of the test. While the feedback varies greatly across the activities the most common pattern is the contrast between the expectation and the actual state of the submitted solution. In most of the cases, the focus of the auto-grader is on the correctness of the solution. However, there are several activities focusing on code style and quality. In those activities, the grader goes beyond correctness and evaluates the compliance of the submitted solution. Figure \ref{fig:auto-grader} shows an example interaction with the auto-grader, illustrating how the feedback facilitates iterative improvement of the solution.

\section{Experimental Design}
\subsection{Models}
The original GPT model \cite{radford2018improving} is a 12-layer decoder-only transformer \cite{vaswani2017attention} with masked self-attention heads. Its core capability is fine-tuning on a downstream task. The GPT-2 model~\cite{radford2019language} largely follows the details of the original GPT model with a few modifications, such as layer normalization moved to the input of each sub-block, additional layer-normalization after the first self-attention block, and a modified initialization. Compared to the original model it displays remarkable multi-task learning capabilities~\cite{radford2019language}. The third generation of GPT models~\cite{brown2020language} uses almost the same architecture as GPT-2. The only difference is that it alternates dense and locally banded sparse attention patterns in the layers of the transformer. The main focus of Brown et al. was to study the dependence of performance and model size where eight differently sized models were trained (from 125 million to 175 billion parameters). The largest of the models is commonly referred to as GPT-3. The interesting property of these models is that they appear to be very strong zero- and few-shot learners. This ability appears to improve with the increasing size of the model~\cite{brown2020language}. The technical details about the recently released GPT-4 model have not been disclosed due to (alleged) concerns about potential misuses of the technology as well as highly competitive market with generative AI \cite{openai2023gpt4}.

We are primarily interested in the performance of the \verb|gpt-4| (GPT-4) model as compared to \verb|text-davinci-003| (GPT-3.5). As of writing of this paper, GPT-4 is by far the most advanced model released by OpenAI. The model is focused on dialog between a user and a system. On the other hand, GPT-3.5 is a more general model focused on text completion. It builds on top of previous \verb|text-davinci-002|, which in turn is based on \verb|code-davinci-002| (focused on code-completion tasks) which is sometimes referred to as codex. To gauge the rate of improvement over the several recent years, we compare the performance of GPT-4 to GPT-3.5 as well as to the previous generation's InstructGPT \verb|text-davinci-001| model (GPT-3)\footnote{OpenAI: Model index for researchers. Available at: \url{https://beta.openai.com/docs/model-index-for-researchers/instructgpt-models} [Accessed 2023-01-15]} on the MCQ answering task. For coding exercises, we benchmark GPT-4 to GPT-3.5 only. This is because GPT-3 is mostly focused on text completion, and is not capable of producing (decent) solutions to coding exercises; this ability only emerged with \verb|code-davinci-002| and later models.

We set the \verb|temperature| of all the models to 0.0, which corresponds to no randomness. The higher the \verb|temperature| the more creative the output but it can also be less factual. As the temperature approaches 0.0, the model becomes more deterministic, which we deem as important for reproducability. Given that existing literature does use different temperatures for testing, we did initially test a variety of temperatures, but found that setting temperature to 0.0 worked well for our setting, which falls inline with the findings and precedence of existing work regarding multiple-choice questions  \cite{bommarito2023gpt, Lievin2022CanLL, pan2022}. Additionally, given that we were largely evaluating questions automatically a single-time per question, setting temperature to 0.0 provided us with the most likely completion of GPT-4, allowing us to be more confident in our resulting analysis. We set \verb|max_tokens| to 500 (a token roughly corresponds to a word) for MCQ answering, and to 2,000 (GPT-3.5) or 4,000 (GPT-4) for coding activities. This parameter controls the maximum length of the completion (i.e., the output). Note that each model has a length limit on the prompt, and the completion counts towards that limit. While GPT-4 allows for 8,192 tokens\footnote{There is also a variant of the model that supports up to 32,768 tokens.} the GPT-3.5 can only accept up to 4,097 tokens. We set \verb|top_p| to 1, as is recommended  when \verb|temperature| is set to 0.0. This parameter is related to \verb|temperature| and also influences creativeness of the output. We set \verb|frequency_penalty| to 0, which allows repetition by ensuring no penalty is applied to repetitions. Finally, we set \verb|presence_penalty| to 0, ensuring no penalty is applied to tokens appearing multiple times in the output.

\subsection{Experimental Design}
To test the performance on MCQs, we submit questions one by one using the \verb|openai| Python library\footnote{GitHub: OpenAI Python Library. Available at: \url{https://github.com/openai/openai-python} [Accessed 2023-01-16]} which is a wrapper for the OpenAI's REST API. For GPT-3 and GPT-3.5, we embed each question in the prompt templates shown in Figure \ref{fig:mcq-prompt-template}. Since GPT-4 is a model optimized for dialogue, we use different prompts---the ones shown in Figure \ref{fig:mcq-prompt-template-gpt4}. Note that the prompt for GPT-4 is designed with the intent to prevent the model from explaining the answer to a user as we are only interested in the answer(s) themselves. Each model returns one or more of the choices as the prompt completion (response), which is then compared to the reference answer. Following the approach adopted by PE1 and PE2, partially correct answers are considered to be incorrect.

\begin{figure}
\footnotesize
\begin{Verbatim}[frame=single,commandchars=\\\{\}]
I am a highly intelligent bot that can easily handle answering 
multiple-choice questions on introductory Python topics. 
Given a question and choices I can always pick the right ones.

Question: \textcolor{blue}{\string{\string{question\string}\string}}

Choices:
\textcolor{blue}{\string{\string{choices\string}\string}}

The correct answer:
\end{Verbatim}
\begin{textblock*}{3.4cm}(6.3cm,-2.7cm)
\circled{1}
\end{textblock*}
\begin{textblock*}{3.4cm}(1.5cm,-1.95cm)
\circled{2}
\end{textblock*}
\begin{textblock*}{3.4cm}(0.1cm,-1.1cm)
\circled{3}
\end{textblock*}
\caption{MCQ Prompt Template for GPT-3 and GPT-3.5. The text of the preamble (1) is inspired by OpenAI's QA example. The \string{\string{question\string}\string} token~(2) is replaced with the question text. The \string{\string{choices\string}\string} token~(3) is replaced with the candidate answers where each one is placed on a single line preceded by a capital letter.}
\label{fig:mcq-prompt-template}
\end{figure}

\begin{figure}
\footnotesize
\begin{Verbatim}[frame=single,commandchars=\\\{\}]
You are a highly intelligent bot that can easily handle answering
multiple-choice questions on introductory Python topics. Given a 
question and choices you can always pick the right ones. You are 
not expected to explain the answers.

Example user question:
What function in Python is typically used to display text to the 
terminal?
A. input
B. print
C. len
D. int

Example bot response:
B. print
\end{Verbatim}

\begin{textblock*}{3.4cm}(4.8cm,-1.7cm)
\begin{Verbatim}[frame=single,commandchars=\\\{\}]
Question: \textcolor{blue}{\string{\string{question\string}\string}}

Choices:
\textcolor{blue}{\string{\string{choices\string}\string}}
\end{Verbatim}
\end{textblock*}

\begin{textblock*}{3.4cm}(3.2cm,-3.5cm)
\circled{1}
\end{textblock*}
\begin{textblock*}{3.4cm}(0cm,-1.55cm)
\circled{2}
\end{textblock*}
\begin{textblock*}{3.4cm}(6.2cm,-1.6cm)
\circled{3}
\end{textblock*}
\begin{textblock*}{3.4cm}(4.85cm,-.75cm)
\circled{4}
\end{textblock*}
\caption{MCQ Prompt Templates for GPT-4. The outer frame shows the system's prompt which is used to set the context of the dialogue. The text of the preamble (1) is inspired by OpenAI's QA example. The example user question and bot response (2) primes the model to return the answer(s) only (no explanations). The inner frame depicts the user's message sent to the dialogue system. The \string{\string{question\string}\string} token~(3) is replaced with the question text. The \string{\string{choices\string}\string} token~(4) is replaced with the candidate answers where each one is placed on a single line preceded by a capital letter.}
\label{fig:mcq-prompt-template-gpt4}
\end{figure}

In coding tasks, we submit the instructions using the prompt templates shown in Figure \ref{fig:project-prompt-template} for GPT-3.5. Again, we use a different prompt for GPT-4 which is shown in Figure \ref{fig:project-prompt-template-gpt4}. While we needed to embed the coding activity instructions into the GPT-3.5's prompt (as shown in Figure \ref{fig:project-prompt-template}) these are passed to GPT-4 more naturally as a message coming from a user.

\begin{figure}
\footnotesize
\begin{Verbatim}[frame=single,commandchars=\\\{\}]
TASK
Implement a Python program to print "Hello, World!" in hello.py.
=== hello.py ===
# TODO 1
===

SOLUTION
=== hello.py ===
print("Hello, World!")
===

TASK
\textcolor{blue}{\string{\string{instructions\string}\string}}
=== \textcolor{blue}{\string{\string{file_name\string}\string}} ===
\textcolor{blue}{\string{\string{handout\string}\string}}
===

SOLUTION
\end{Verbatim}

\begin{textblock*}{3.4cm}(11.4cm,-3.7cm)
\begin{Verbatim}[frame=single,commandchars=\\\{\}]
[...]

SOLUTION
=== \textcolor{blue}{\string{\string{file_name\string}\string}} ===
\textcolor{blue}{\string{\string{solution\string}\string}}
===

FEEDBACK
\textcolor{blue}{\string{\string{feedback\string}\string}}

FIXED SOLUTION
\end{Verbatim}
\end{textblock*}
\begin{textblock*}{1.4cm}(1.5cm,-4.05cm)
\circled{1}
\end{textblock*}
\begin{textblock*}{1.4cm}(1.7cm,-1.95cm)
\circled{2}
\end{textblock*}
\begin{textblock*}{1.4cm}(2.3cm,-1.66cm)
\circled{3}
\end{textblock*}
\begin{textblock*}{1.4cm}(1.06cm,-1.33cm)
\circled{4}
\end{textblock*}
\begin{textblock*}{1.4cm}(13.7cm,-2.75cm)
\circled{5}
\end{textblock*}
\begin{textblock*}{1.4cm}(12.55cm,-2.45cm)
\circled{6}
\end{textblock*}
\begin{textblock*}{1.4cm}(12.6cm,-1.33cm)
\circled{7}
\end{textblock*}

\caption{Coding Task Prompt Templates for GPT-3.5. Outer frame is the first submission template. The preamble (1) primes the model to generate a solution code as completion. The \string{\string{instructions\string}\string} token~(2) is replaced with the coding task instructions. A starter code is injected into \string{\string{file\_name\string}\string} (3) and \string{\string{handout\string}\string} (4) tokens. The inner frame shows the template for re-submission (appended to the original). The \string{\string{file\_name\string}\string}~(5) and \string{\string{solution\string}\string} (6) were replaced with the GPT's solution and the \string{\string{feedback\string}\string} (7) with the auto-grader's feedback.}
\label{fig:project-prompt-template}
\end{figure}

\begin{figure}
\footnotesize
\begin{Verbatim}[frame=single,commandchars=\\\{\}]
You are a highly intelligent coding bot that can easily handle any 
Python programming task. Given a natural language instructions you 
can always implement the correct Python solution. Your focus is the 
solution code only. You are not allowed to provide explanations.

Example (toy) instructions:
Implement a Python program to print "Hello, World!" in the 
hello.py.

Example bot solution:
=== hello.py ===
print("Hello, World!")
===
\end{Verbatim}

\begin{textblock*}{1.4cm}(3.6cm,-2.7cm)
\circled{1}
\end{textblock*}
\begin{textblock*}{1.4cm}(6.9cm,-2.2cm)
\circled{2}
\end{textblock*}
\begin{textblock*}{1.4cm}(2.4cm,-1.1cm)
\circled{3}
\end{textblock*}

\caption{Coding Task Prompt Template for GPT-4. The preamble (1) primes the model to generate the code of the solution only (no explanations). The example instructions (2) and solution (3) are used to further clarify the expectations on the output.}
\label{fig:project-prompt-template-gpt4}
\end{figure}

To each submission, the auto-grader assigns a score and generates detailed actionable feedback. If the full score was not achieved we amended the GPT-3.5's prompt with the addendum (shown in Figure~\ref{fig:project-prompt-template}). For GPT-4 we simply continued in the dialogue where the solution it generated was followed by the auto-grader's feedback. Then, we submitted the revised solution to the auto-grader and repeated the process until either the full score was achieved or the solution remained unchanged from the preceding one (impasse).

When dealing with coding activities, we encountered the models' limitation on the prompt length (4,097 tokens for GPT-3.5 and 8,192 or 32,768 for GPT-4). Within this limit, it was necessary to fit: (i) the prompt boilerplate; (ii) the instructions; (iii) the contents of the handout files (usually starter code) distributed to learners; and (iv) the solution generated by the model (i.e., the prompt completion). Instead of full project (8) instructions we submitted the individual project tasks (32). If a task could not be fitted into a prompt, we decreased the \verb|max_tokens| parameter (space for solution) to <2,000 for GPT-3.5 or <4,000 for GPT-4. If this did not resolve the issue we edited the instructions leaving out pieces that could be reasonably expected as not being useful for the GPT models. As the last resort, we would split the task into several smaller coding activities (69 overall) if the task was to develop several loosely coupled elements. The GPT models' solutions were then submitted to the auto-grader.

\section{Results and Discussion}

\subsection{(RQ1) To what degree can GPT-4 generate correct answers to MCQs?}
The results of applying the GPT models to the MCQ exercises are reported in Tables \ref{tab:pe1_results} (PE1), \ref{tab:pe2_results} (PE2), and \ref{tab:ppp_concepts_results} (PE3). While the original GPT-3 model correctly answered only 199 out of the 530 questions (37.5\%), the GPT-3.5 and GPT-4 models were much more successful. GPT-3.5 correctly answered 341 MCQs (64.3\%). GPT-4 successfully handled 446 questions (84.1\%). Hence, we observe a sizeable improvements across the successive generations of the GPT models.

The results from PE1 are reported in Table \ref{tab:pe1_results}. In order to pass the course the score of 70\% or better is required from all the 5 tests. While the GPT-3 model could not pass any of the tests, and the more successful GPT-3.5 model passed only the first course module, as already reported in prior work \cite{savelka2023large,Savelka2023}, the GPT-4 model passed all the four module tests as well as the summary test (i.e., passing the course with the overall score of 85\%).

\begin{table*}
  \caption{PE1 results. The graded assignments are colored; green and check mark indicate passing while red means failing.}
  \label{tab:pe1_results}
  \setlength{\tabcolsep}{4pt}
  \begin{tabular}{l|rrr|rrr}
    \toprule
                 &\multicolumn{3}{c|}{Quizzes} &\multicolumn{3}{c}{Tests} \\
    Module Topic                             &\multicolumn{1}{c}{GPT-3}&\multicolumn{1}{c}{GPT-3.5}&\multicolumn{1}{c}{GPT-4}&\multicolumn{1}{|c}{GPT-3}&\multicolumn{1}{c}{GPT-3.5}&\multicolumn{1}{c}{GPT-4}\\
    \hline
    Introduction to Python and programming   &8/10 (80.0\%)&10/10 (100\%)&10/10 (100\%)&\color{Red}6/10 (60.0\%)&\color{OliveGreen}\checkmark\ 9/10 (90.0\%)&\color{OliveGreen}\checkmark\ 10/10 (100\%)\\
    Data types, variables, I/O, operators    &6/10 (60.0\%)&10/10 (100\%)&10/10 (100\%) &\color{Red}6/20 (30.0\%)&\color{Red}10/20 (50.0\%)&\color{OliveGreen}\checkmark\ 18/20 (90.0\%)\\
    Booleans, conditionals, loops, operators &3/10 (30.0\%)&7/10 (70.0\%)&10/10 (100\%) &\color{Red}6/20 (30.0\%)&\color{Red}12/20 (60.0\%)&\color{OliveGreen}\checkmark\ 16/20 (80.0\%)\\
    Functions, data structures, exceptions   &6/12 (50.0\%)&9/12 (75.0\%)&9/12  (75.0\%)&\color{Red}7/22 (31.8\%)&\color{Red}12/22 (54.5\%)&\color{OliveGreen}\checkmark\ 20/22 (90.9\%)\\
    Completion (Summary Test)            &-            &-            &-                 &\color{Red}7/35 (20.0\%)&\color{Red}17/35 (48.6\%)&\color{OliveGreen}\checkmark\ 27/35 (77.1\%)\\
    \hline
    \bf Course Total                     &\bf 23/42    &\bf 36/42    &\bf 39/42    &\bf 32/107   &\bf 60/107   &\bf 91/107        \\
                                         &\bf (54.8\%) &\bf (85.7\%) &\bf (92.4\%) &\bf (29.9\%) &\bf (56.1\%) &\bf (85.0\%)      \\
    \bottomrule
  \end{tabular}
\end{table*}

The performance of the models on PE2 is presented in Table \ref{tab:pe2_results}. The assessment scheme of PE2 is the same as that of PE1. Here, the GPT-3.5 model was somewhat more successful and passed 3/4 module tests. However, it also failed the Summary Test (65.0\%). The original GPT-3 model still could not pass a single test. These findings were also reported in prior work \cite{savelka2023large,Savelka2023} The GPT-4 model again passed all the five graded assignments (89.6\% overall score).

\begin{table*}
  \caption{PE2 results. The graded assignments are colored; green and check mark indicates passing while red means failing.}
  \label{tab:pe2_results}
  \setlength{\tabcolsep}{4pt}
  \begin{tabular}{l|rrr|rrr}
    \toprule
                 &\multicolumn{3}{c|}{Quizzes} &\multicolumn{3}{c}{Tests} \\
    Module Topic                             &\multicolumn{1}{c}{GPT-3}&\multicolumn{1}{c}{GPT-3.5}&\multicolumn{1}{c}{GPT-4}&\multicolumn{1}{|c}{GPT-3}&\multicolumn{1}{c}{GPT-3.5}&\multicolumn{1}{c}{GPT-4}\\
    \hline
    Modules, packages, and PIP                     &3/10 (30.0\%)&6/10 (60.0\%)&10/10 (100\%) &\color{Red}10/18 (55.6\%)&\color{OliveGreen}\checkmark\ 14/18 (77.8\%)&\color{OliveGreen}\checkmark\ 17/18 (94.4\%)\\
    Strings, string list methods, exceptions       &7/10 (70.0\%)&6/10 (60.0\%)&9/10 (90.0\%) &\color{Red}4/15 (26.7\%) &\color{OliveGreen}\checkmark\ 11/15 (73.3\%)&\color{OliveGreen}\checkmark\ 13/15 (86.7\%)\\
    Object-oriented programming                    &7/10 (70.0\%)&8/10 (80.0\%)&9/10 (90.0\%) &\color{Red}4/17 (23.5\%) &\color{OliveGreen}\checkmark\ 12/17 (70.6\%)&\color{OliveGreen}\checkmark\ 15/17 (82.4\%)\\
    Miscellaneous                                  &8/12 (66.7\%)&9/12 (75.0\%)&11/12 (91.7\%)&\color{Red}4/16 (25.0\%) &\color{Red}9/16 (56.2\%)  &\color{OliveGreen}\checkmark\ 15/16 (93.8\%)\\
    Completion (Summary Test)                      &-            &-            &-             &\color{Red}11/40 (27.5\%)&\color{Red}26/40 (65.0\%) &\color{OliveGreen}\checkmark\ 35/40 (87.5\%)\\
    \hline
    \bf Course Total                               &\bf 25/42    &\bf 29/42    &\bf 40/42   &\bf 33/106                   &\bf 72/106                   &\bf 95/106 \\
                                                   &\bf (59.5\%) &\bf (69.0\%) &\bf (95.2\%)&\bf (31.1\%)                 &\bf (67.9\%)                 &\bf (89.6\%) \\
    \bottomrule
  \end{tabular}
\end{table*} 

Table \ref{tab:ppp_concepts_results} reports the results of applying the models to the MCQs in PPP. Again, we observe similar progression from the weakest GPT-3 model (30.9\% on the tests), through the better performing GPT-3.5 (65.4\%), as already reported in prior work \cite{savelka2023large,Savelka2023}, to the best performing GPT-4 model (77.8\%). Note, that passing of PPP is not solely determined by the MCQ assessments, and largely depends on the performance on the coding activities (projects).

\begin{table*}
  \caption{PPP results. The tests contribute to the grade, typically by no more than 20\%. Since in PPP tests themselves do not determine pass or fail no colors are used.}
  \label{tab:ppp_concepts_results}
  \setlength{\tabcolsep}{4pt}
  \begin{tabular}{l|rrr|rrr}
    \toprule
                                                &\multicolumn{3}{c|}{Quizzes} &\multicolumn{3}{c}{Tests} \\
    Module Topic                                & GPT-3        & GPT-3.5      & GPT-4        & GPT-3      & GPT-3.5    & GPT-4     \\
    \hline
    Python basics and introduction to functions &12/30 (40.0\%)&21/30 (70.0\%)&27/30 (90.0\%)&4/12 (33.3\%)&9/12 (75.0\%) &10/12 (83.3\%)\\
    Control flow, strings, input and output     &8/22 (36.4\%) &10/22 (45.5\%)&16/22 (72.7\%)&3/11 (27.3\%)&8/11 (72.7\%) &10/11 (90.9\%)\\
    Python data structures                      &9/18 (50.0\%) &10/18 (55.6\%)&14/18 (77.8\%)&4/14 (28.6\%)&9/14 (64.3\%) &10/14 (71.4\%)\\
    Object-oriented programming                 &6/14 (42.9\%) &7/14 (50.0\%) &11/14 (77.6\%)&4/11 (36.4\%)&10/11 (90.9\%)&11/11 (100\%)\\
    Software development                        &9/19 (47.4\%) &12/19 (63.2\%)&16/19 (84.2\%)&5/10 (50.0\%)&7/10 (70.0\%) &10/10 (100\%)\\
    Data manipulation                           &6/17 (35.3\%) &9/17 (52.9\%) &13/17 (76.5\%)&5/13 (38.5\%)&5/13 (35.5\%) &8/13 (61.5\%)\\
    Web scraping and office document processing &5/10 (50.0\%) &5/10 (50.0\%) &5/10 (50.0\%) &0/5 (0.0\%)  &3/5 (60.0\%)  &3/5 (60.0\%)\\
    Data analysis                               &6/22 (27.3\%) &17/22 (77.3\%)&18/22 (81.8\%)&0/5 (0.0\%)  &2/5 (40.0\%)  &2/5 (20.0\%)\\
    \hline
    \bf Course Total                            &\bf 61/152   &\bf 91/152   &\bf 120/152 &\bf 25/81   &\bf 53/81   &\bf 64/81\\
                                                &\bf (40.1\%) &\bf (59.9\%) &\bf (78.9\%)&\bf (30.9\%)&\bf (65.4\%)&\bf (79.0\%) \\
    \bottomrule
  \end{tabular}
\end{table*}

Overall, the findings reported in prior work \cite{savelka2023large,Savelka2023} no longer hold. While GPT-3 and GPT-3.5 models' performance on the programming MCQs is not sufficient for passing the three courses, GPT-4 handles the MCQs well enough to reliably pass the course MCQ assessments. Note, that in some countries much lower passing scores may be required. Hence, our finding of GPT-4 passing the assessments the prior models fail to pass might not hold in those contexts.

\subsection{(RQ2) Does GPT-4 struggle with programming MCQs containing code?}
Table \ref{tab:results} reports the results of our experiments on how GPT models handle MCQs of various types. The GPT-4 model performs the best (84.5\% overall) with quite a noticeable margin over the GPT-3.5 (65.5\% overall). The performance of the original GPT-3 appears to be much lower compared to the other two models. This is to be expected, as the major breakthrough in OpenAI GPT models' capabilities in handling computer code was Codex (\verb|code-davinci-002|) \cite{https://doi.org/10.48550/arxiv.2107.03374} which is the predecessor of \verb|text-davinci-003| (i.e., GPT-3.5).\footnote{OpenAI: Model index for researchers. Available at: \url{https://beta.openai.com/docs/model-index-for-researchers/instructgpt-models} [Accessed 2023-01-15]}

\begin{table}
  \caption{Performance of the GPT models across MCQs of different types.}
  \label{tab:results}
  \centering
  \begin{tabular}{lrrr}
  \hline
    Question Type & GPT-3  & GPT-3.5     & GPT-4 \\
  \hline
    \multicolumn{4}{c}{\bf No Code}      \\
  \hline
    True/False                    & 13/25      & 20/25      & 23/25   \\
                                  & (52.0\%)   & (80.0\%)   & (92.0\%)\\
    Identify True/False Statement & 12/44      & 27/44      & 35/44   \\
                                  & (27.3\%)   & (61.4\%)   & (79.5\%)\\
    Finish Statement              & 42/56      & 50/56      & 56/56 \\
                                  & (75.0\%)   & (89.3\%)   & (1.0\%)\\
    Other                         & 25/47      & 37/47      & 43/47   \\
                                  & (53.2\%)   & (78.7\%)   & (91.5\%)\\
    \bf Total                     &\bf 92/172  &\bf 134/172 &\bf 157/172   \\
                                  &\bf (53.5\%)&\bf (77.9\%)&\bf (91.3\%)\\
  \hline
    \multicolumn{4}{c}{\bf With Code}      \\
  \hline
    True/False                    & 12/23      & 10/23      & 13/23 \\
                                  & (52.2\%)   & (43.5\%)   & (56.5\%)\\
    Identify True/False Statement & 10/26      & 11/26      & 16/26 \\
                                  & (38.5\%)   & (42.3\%)   & (61.5\%)\\
    Output                        & 28/110     & 53/110     & 86/110\\
                                  & (25.4\%)   & (48.2\%)   & (78.2\%)\\
    Fill-in                       & 5/13       & 11/13      & 12/13\\
                                  & (38.5\%)   & (84.6\%)   & (92.3\%)\\
    Finish Statement              & 10/27      & 22/27      & 25/27\\
                                  & (37.0\%)   & (81.5\%)   & (92.6\%)\\
    Other                         & 42/159     & 106/159    & 139/159\\
                                  & (26.4\%)   & (66.7\%)   & (87.4\%)\\
    \bf Total                     &\bf 107/358 &\bf 213/358 &\bf 291/358   \\
                                  &\bf (29.9\%)&\bf (59.5\%)&\bf (81.3\%)\\
  \hline
    Overall                       & 199/530    & 347/530    & 448/530   \\
                                  & (37.5\%)   & (65.5\%)   & (84.5\%)\\
  \hline
  \end{tabular}
\end{table}

There appears to be a clear difference between the performance of the most capable GPT-4 on the MCQs that contain code snippets (81.0\% overall) compared to those that do not (90.7\% overall). This is to be expected as the combination of code and natural language likely constitutes (on average) more complex input than natural language alone. Additionally, it is quite possible that in our particular context the questions with code are (on average) more difficult than questions with no code. However, notice that the gap appears to be much wider in the preceding generations of the GPT models (29.9\% vs 53.3\% for GPT-3 and 59.5\% vs 77.9\% for GPT-3.5). Hence, it appears that GPT-4's capabilities in handling MCQs with code are much improved compared to its predecessors. However, the observable difference between the performance on MCQs with natural language only and MCQs with code remains. There also appears to be clear difference between the performance of GPT-4 on the completion-oriented MCQs (96.9\%), i.e., \emph{Finish Statement} and \emph{Fill-in} and the rest (81.3\%). Since GPT models are primarily focused on prompt completion, be it text or computer code, this finding is also as expected. Hence, the findings from prior work \cite{savelka2023large} still hold in this regard.

To investigate further GPT-4's code handling limitations, we analyzed the 67 MCQs with code that GPT-4 answered incorrectly, manually inspecting the full answers, and sometimes altering the prompt and requerying, to hypothesize the reasons for the errors.  We found that the model's mistakes fell into five main categories, listed in Table~\ref{tab:mcq_code_qualcodes}.

\begin{table}
  \caption{Qualitative coding of wrong GPT-4 code MCQ answers}
  \label{tab:mcq_code_qualcodes}
  \centering
  \begin{tabular}{p{3cm}p{4cm}r}
  \hline
  Code & Definition & \#\ Q \\
\hline
WRONG-Q & GPT's explanation revealed a valid interpretation suggesting that the question itself was ambiguous & 13 \\ 
MULTI-HOP & Model was confused by multi-hop reasoning. & 19 \\ 
FLUKE & GPT only gets the question wrong when temperature=0 & 8 \\ 
DUAL-INTERPRETATION &  The text of the code did not match its intent, and GPT sometimes relied on the intent & 13 \\ 
INCOHERENT & GPT answers are variable and inconsistent & 14 \\ 
\hline
\end{tabular}
\end{table}

\textbf{Problems in the question (WRONG-Q)} GPT's explanations in some cases exposed problems in the original questions.  For example, one question 
asks, in part, ``Which of the potential solutions would load the data into a \texttt{DataFrame} object where the index is set to the \texttt{month} column?''. The correct answer was ``\texttt{df = pd.read\_csv("tickets\_monthly.csv", index\_col=0)}'', using the Pandas package to load a CSV file into a variable \texttt{df}.  However GPT-4, answered ``\texttt{pd.read\_csv( "tickets\_monthly.csv", index\_col=0)}'', explaining that the actual correct answer ``... is correct in terms of functionality, but it assigns the \texttt{DataFrame} to a variable \texttt{df}. The question asked for a solution that loads the data into a \texttt{DataFrame} object, not a variable assignment.'' Note that the answer preferred byt the GPT-4 model would be defensible since it does in fact load a \texttt{DataFrame}, albeit in an impractical way. Realistically, we might have granted a student the point if they provided similar explanation . Also note that GPT-4's explanation is not completely correct. The correct answer \emph{does} load a \texttt{DataFrame} as well as assign it. It is then the better answer of the two (because more pragmatic).

\textbf{Multi-hop reasoning (MULTI-HOP)}
GPT is known to have difficulty with answering questions that require multiple hops of reasoning \cite{Wei2022ChainOT}.  Adding an instruction to show the reasoning steps sometimes improves the answers, or even leads to revising a wrong answer after explaining it.  We re-queried GPT4 for some questions that appeared to use multi-hop reasoning, changing the prompt to encourage explanation (we removed the prompt text discouraging explanations, and added, ``Give your answer, then provide your reasoning:'' at the end of the prompt.)  This indeed corrected some, but not all, of the errors. Interestingly, even when GPT-4 lists out the steps, its analysis of a step may be incorrect in the context of a complex, multi-hop answer, even though it is correct when answering a similar question in isolation. Sometimes this mimics \textit{motivated reasoning}.  For example, one question involved predicting the effect of two apparently opposite string replacement commands: 
\begin{verbatim}
quote = '"The things that make me different are the ' +\
        'things that make me, me."'
new_sentence = quote.replace(" different", ", me")
new_sentence = new_sentence.replace(", me",
                                    " different")
\end{verbatim}
Although the second replacement does not restore the original sentence, GPT's step claims that it does:
\begin{quote}
   The second \texttt{replace()} function call replaces the first occurrence of ", me" with " different". 
   The new string becomes: "The things that make me different are the things that make me, me."
\end{quote}
This is not true as the call to the \texttt{replace} function replaces all matching strings. GPT-4 explains this aspect of the question correctly when asked in isolation: 

\begin{quote}
This code will output a modified version of the quote string where every occurrence of the substring ``, me'' is replaced with the string `` different''. [...] ``The things that make me different are the things that make me different.''
\end{quote}

Another commonly occurring failure due to multi-hop reasoning seems to be related to interference between subparts of a question.  For example, one question defines a function called \texttt{mystery} that does a series of edits and permutations on a list. The choices list four calls to this function, each with a proposed output, and asks the student to identify which of the pairs is correct.  Unlike the true ``multi-hop'' questions, these do not build on each other, but they nonetheless seem to interact. GPT-4 can explain the function in isolation and identify the correct result. However, when the four choices are posed together for verification, it identifies all four of them as correct, even though only one is in fact correct.

\textbf{Unlikely failure (FLUKE)}
For MCQs, we set \verb|temperature| to 0 as we were explicitly focused on running a reproducible experiment, capturing the performance of the most likely output of the model. Setting a higher \verb|temperature| and taking the most common of several answers could be a way to achieve more reliably correct performance.  We marked as FLUKE any questions that GPT-4 got right with \verb|temperature| being set higher. This is because in this case the incorrect handling is associated with a particular parametrization of the model.


\textbf{Reasoning biased by inferred code intent (DUAL-INTERP)}
GPT-4 sometimes provided answers that focused on the intent rather than the exact nature of the code itself. For example, in one question, a calculation is performed but the last line, printing out the result, is commented out.  GPT-4 answered as if the line were not commented out. If asked to explain the reasoning step-by-step, GPT-4 sometimes caught its mistake.
GPT-4 is optimized for dialog, in which humans do the best to make sense of inconsistent inputs; we infer the most plausible coherent interpretation of an interlocutor.  A debugging-focused question that asks about unintended behavior from subtly wrong code is interpreted as if the code were ``correct''. The same robust ability to interpret intent that lets it answer poorly-worded English questions apparently trips it up in questions about purposefully misleading code.

\textbf{Inconsistent reasoning (INCOHERENT)}
GPT-4 sometimes gives different answers when queried repeatedly with \verb|temperature| > 0. When providing reasons, it may give contradictory answers within the same response.  For example, one question asked if the expression \texttt{not (mag < 5)} would be correct in a program with multiple blanks, at a point where values >= 6 have already been ruled out.  GPT-4 incorrectly responds ``The suggested expression \texttt{not (mag < 5)} is not accurate because it will also include magnitudes of 6.0 and higher.'' However, it then gives the completed code, using the logically equivalent expression \texttt{mag > 5.0}. It explains that ``should be mag >= 5.0, as this accurately identifies the \verb|"Moderate"| category without including higher magnitude levels.''  Other responses contained similar contradictions.  



\subsection{(RQ3) To what degree can GPT-4 produce solutions to complex coding tasks?}
The results of applying the GPT models to the coding activities are reported in Table \ref{tab:project_results}. While the GPT-3.5 model obtained 407 points from the available 760 (53.6\%) after a single submission to each activity GPT-4 collected 545 points (71.7\%). While there is no stable grading scheme for PPP it is fair to anticipate that the performance of GPT-3.5 on MCQ tests (65.4\%) and projects (54.6\%) would typically not be deemed sufficient for passing the course. On the other hand, the performance of GPT-4, i.e., 79\% on the MCQ tests and 71.7\% on the projects, comes dangerously close if not all the way towards actually passing the course. Hence, the findings reported in \cite{Savelka2023} appear to be challenged by the more capable GPT-4 model as well as the findings reported in \cite{10.1145/3511861.3511863,finnie2023my,denny2023conversing,piccolo2023many}.

We observe that the performance of GPT-4 across the tasks appears to be related to the performance of GPT-3.5. That is to say, the tasks that were challenging for GPT-3.5 appear to be challenging even for GPT-4. GPT-3.5 collected very low scores on the first submission from projects 2 (40 points), 5 (4 points), 7 (20 points), and 8 (11 points). While GPT-4 obtained somewhat higher scores (56, 14, 53 and 44 points respectively) these four projects remained the most challenging. Hence, it appears that the workings of GPT-3.5 and GPT-4 with respect to the coding tasks are not fundamentally different, despite GPT-4 being noticeably better performing. Hence, the strengths and limitations reported in prior work \cite{Savelka2023,10.1145/3511861.3511863,finnie2023my,denny2023conversing,piccolo2023many} still hold to a certain degree.

\begin{table*}
  \caption{Coding tasks results. Max score is the maximum score achievable from a task. First score is the score after first submission. Resubs is the number of re-submissions after the first submission before the full score or no-change impasse were reached. Final score is the score after feedback.}
  \label{tab:project_results}
  \setlength{\tabcolsep}{3.5pt}
  \begin{tabular}{llr|rrr|rrr}
    \toprule
                  &               &    &\multicolumn{3}{c|}{GPT-3.5}&\multicolumn{3}{c}{GPT-4}\\
    Project Topic &Tasks (skills) &Max &1st &Resubs &Final &1st &Resubs &Final\\
    \midrule
    Types, variables, functions &Variable assignment                               &13 &13 &0 &13 &13 &0 &13\\
                                &User-defined functions, imports, return vs print  &20 &16 &1 &16 &20 &0 &20\\
                                &Simple scripts, comments                          &50 &50 &0 &50 &50 &0 &50\\
                                &Testing                                           &12 &12 &0 &12 &12 &0 &12\\
                                &\bf Project 1 Total                               &\bf 95 &\bf 91 &\bf 1 &\bf 91 &\bf 95 &\bf 0 &\bf 95\\
    \midrule
    Iteration, conditionals,    &Conditional statements                            &10 &10 &0 &10 &10 &0 &10\\
    \ strings, basic I/O        &Strings, while loop, for loop, complex printing   &35 &5 &4 &9 &6 &11&18\\
                                &Read and write files                              &15 &0  &2 &3  &15 &0 &15\\
                                &Complex script                                    &35 &25 &1 &25 &25 &1 &25\\
                                &\bf Project 2 Total                               &\bf 95 &\bf 40 &\bf 7 &\bf 47 &\bf56 &\bf 12&\bf 68 \\
    \midrule
    Lists, sets, tuples and     &Create container, access, add, remove and update  &40 &38 &1 &40 &40 &0 &40\\
    \ dictionaries              &\ elements, and convert containers across types   &   &   &  &\\
                                &Nested data structure, transformation, sorting,   &55 &30 &3 &40 &55 &0 &55 \\
                                &\ export to file, complex report                  &   &   &  &\\
                                &\bf Project 3 Total                               &\bf 95    &\bf 68 &\bf4 &\bf 80 &\bf 95 &\bf0 &\bf 95 \\
    \midrule
    Classes, objects,           &Implement classes                                 &13 &13 &0 &13 &13 &0 &13\\
    \ attributes and methods    &Define, access and set private attributes         &19 &19 &0 &19 &19 &0 &19\\
                                &Inheritance                                       &20 &20 &0 &20 &20 &0 &20\\
                                &Implement re-usable utility functions             &16 &16 &0 &16 &16 &0 &16\\
                                &Composition, object instantiation                 &16 &16 &0 &16 &16 &0 &16\\
                                &Override special methods (repr, eq)               &11 &0  &3 &0  &9  &1 &11 \\
                                &\bf Project 4 Total                               &\bf 95 &\bf 84 &\bf 3 &\bf 84 &\bf 93 &\bf 1 &\bf 95\\
    \midrule
    Debugging, refactoring,     &Identify and fix errors in code                   &10 &0 &2  &0  &5  &2 &5 \\
    \ testing and packaging     &Refactor larger code base                         &14 &- &-  &-  &-  &- &- \\
                                &Exception handling                                &13 &4 &2  &4  &4  &2 &13 \\
                                &Analyze and fix code on style correctness         &28 &0 &10 &0  &0  &23&14\\
                                &                                                  &   &  &   &\\
                                &Test-driven development                           &20 &0 &5  &5  &5  &7 &10\\
                                &                                                  &   &  &   &\\
                                &Package Python application using pip              &10  &-   &-   &-  &-   &-   &-   \\
                                &\bf Project 5 Total                               &\bf 95 &\bf 4 &\bf 19 &\bf 9&\bf 14&\bf 34&\bf 42\\
    \midrule
    Files and datastores        &Load and store data in files                      &45 &30 &3 &45 &45 &0 &45\\
                                &Create SQL objects, load and query data in SQL    &30 &30 &0 &30 &30 &0 &30\\
                                &Load and query data in MongoDB                    &20 &20 &0 &20 &20 &0 &20\\
                                &\bf Project 6 Total                               &\bf 95 &\bf 80 &\bf 3 &\bf 95&\bf 95 &\bf 0 &\bf 95 \\
    \midrule
    Web scraping and office     &Get HTML, extract information from HTML,          &35 &5  &10 &18 &33 &1 &35\\
    \ document processing       &\ handle multiple HTML files                      &   &  &   &\\
                                &Manipulate Excel files programatically            &25 &5  &4  &5 &5 &4 &5\\
                                &Authenticate and utilize public API               &15 &0  &1  &0 &15 & 0 &15\\
                                &Manipulate Word files programmatically            &20 &10 &3  &20 & 0 & 1 & 20\\
                                &\bf Project 7 Total                               &\bf 95 &\bf 20 &\bf 18 &\bf 43 &\bf 53 &\bf 6 & \bf 75 \\
    \midrule
    Data analysis               &Load data to pandas, merge pandas DataFrames,         &35 &11 &3 &11  &24 & 2 & 24\\
                                &\ persist pandas DataFrame                            &   &   &  &\\
                                &Assess data quality, examine descriptive statistics   &40 &0  &3 &25  &0  & 1 & 25\\
                                &Utilize regular expressions                           &20 &0  &1 & 20 &20 &0 &20\\
                                &\bf Project 8 Total                                   &\bf 95 &\bf 11  &\bf 7 &\bf 56&\bf 44 &\bf 3 &\bf 69 \\
    \midrule
                                &\bf Course Total                                      &\bf 760 &\bf 407      &\bf 59 &\bf 505 &\bf 545 &\bf  56   &\bf 634\\
                                &                                                      &        &\bf 53.6\% &        &\bf 66.4\% &\bf 71.7\% &    &\bf 83.4\%\\
    \cline{2-9}
  \end{tabular}
\end{table*} 


\subsection{(RQ4) Can GPT-4 successfully utilize feedback to fix solutions of coding tasks?}

The coding tasks results after providing the models with feedback are also reported in Table \ref{tab:project_results}. The overall score achieved by the GPT-3.5 model improved from 53.6\% to 66.4\%. The score would still likely be too low for passing the course. GPT-4's score increased from 71.7\% to 83.4\%. This score would almost certainly enable a human learner to pass the course. Hence, the ability of GPT-4 to utilize feedback seems to be even stronger than the ability of the GPT-3.5 model evaluated in \cite{Savelka2023,piccolo2023many}. That is, of course, if there is no requirement related to passing some minimal threshold for all of the projects. Even after 34 feedback iterations on the project 6 coding activities (debugging, refactoring, testing and packaging) GPT-4 obtained only 42 of the available 95 points. The low performance on this particular project could be explained in terms of several closely related factors. First, certain activities in this project could not be performed at all because they involved use of external tools beyond writing code. For example, the refactoring activities require creating new files and directories as well as renaming and moving files. The packaging activities involve interaction with the command line. Note that it would be possible to equip LLMs, including GPT models, with the ability to manipulate such external tools \cite{mialon2023augmented}. However, we did not consider the possibility in this study, and we have simply refrained from attempting such activities. Similarly, the remaining activities related to fixing errors, style correctness, and testing heavily rely on external tools (i.e., the debugger, \verb|pylint|, \verb|pycodestyle|, \verb|pytest|). These activities were attempted since these tools are not strictly required. However, this is problematic because the feedback would often not contain all the necessary information. For example, in the testing task the feedback contains the high-level information about the test coverage~(\%). The models did not have access to the full coverage report that would show the lines not covered by the tests accessible to human learners.

Some of the coding activities were challenging for GPT-4 because they involved artifacts, beyond the task instructions, that were crucial for generating correct solution. Often, human learners would not necessarily find such tasks particularly challenging. One example of such an artifact is an input data set the size of which exceeds the maximum length of the model's prompt. While human learners can simply inspect the large input data set and identify the appropriate methods to parse the data as required by the task specifications, GPT-4 was unable to solve the task based on the instructions. By extracting sample records from the input data set and providing the sample input along with the expected output, GPT-4 was able to successfully implement the required data transformation.

Finally, we observed that GPT-4, despite being more successful than GPT-3.5, still struggles with fine-grained formatting requirements related to both, the output as well as the code itself. For example, GPT-4 was able to utilize the feedback based on the output of the style-checker that the code contained long lines over 100 characters, and modified the code to shorten the lines. At the same time, the model completely failed to utilize similar feedback from a more strict style-checking tool complaining about the lines that were over 79 characters long. In this particular case, GPT-4 was not able to break up a string that made the line in question 81 characters long. Similarly, when the provided feedback complained about a missing white-space between the \verb;|; character and subsequent number in the task focused on printing a tabular report to a terminal, the GPT-4 model was not able to correct the solution accordingly.






\section{Implications for Teaching Practice}
This study provides evidence that programming instructors need to prepare for a world in which there is an easy-to-use widely accessible technology that can be utilized by learners to collect passing scores, with no effort whatsoever, on what today counts as viable programming knowledge and skills assessments. While this development has been apparent from the growing body of prior work \cite{Savelka2023,10.1145/3511861.3511863,finnie2023my,denny2023conversing,piccolo2023many} this paper is the strongest evidence reported so far in the context of programming education, and it is consistent with the OpenAI's GPT-4 release report \cite{openai2023gpt4}.

In consequence, the instructors may consider shifting the focus from assessment to learning, i.e., they should prioritize the learning experience and skills development, rather than merely preparing learners for assessments. The learners should be encouraged to focus on learning and growth, rather than on always coming up with the right answers. The importance of academic honesty and ethical behavior in the classroom should be emphasized. Ideally, a culture that values original work and personal effort should be promoted. The instructors may need to move away from traditional assessments, such as multiple-choice exams. Instead, they may consider using more complex assessments such as code reviews, pair programming, and oral examinations that require students to demonstrate their understanding in real-time.

While it may be appealing to the instructors to understand the limitations of GPT models when it comes to handling MCQs to design tests that are difficult to be answered automatically we argue that this is likely not a viable approach. Given the rate of improvement over the past several years we document in this study, it appears quite likely the existing limitations will be overcome rather soon, reducing the effectiveness of tests designed to exploit the identified weaknesses. Instead of trying to create ``GPT-proof'' tests, it may be more productive to focus on developing assessments focused on higher-order thinking skills, such as critical thinking, problem-solving, and creativity as these are more difficult for the GPT models to replicate.

In programming activities, we identified several use cases where the application of GPT models does not (yet) appear to be straightforward. While similar argument as the one used for MCQs may be employed against potential hardening of the coding tasks by insisting on unusual fine-grained nuances of the code itself or its output, re-designing the tasks to rely on artifacts beyond the instructions and/or employment of external tools may be promising. This is because the ability to extract, consolidate, and express essential information from multiple sources may become an important skill critical for success of future learners. Instructors may consider incorporating such complex (but not necessary complicated) programming tasks requiring learners to consolidate problem context from multiple sources. This approach could reduce the misuse while promoting beneficial use of GPT models. 

\section{Limitations ant Threats to Validity}
Although, the results of our study provide important insights into the evolving capabilities of the GPT models in passing typical assessments employed in introductory and intermediate Python classes, limitations in several areas must be acknowledged.

\textbf{Generalizability.}
While Python is a widely used and representative language, it is one of many languages used in programming courses at the postsecondary level. GPT models have been shown to handle a number of programming languages \cite{destefanis2023preliminary,bubeck2023sparks}. Nevertheless, our findings may not generalize to those languages with different structures, syntax, and conventions. Secondly, while the programming assessments used in this study are typical of those found in many Python programming courses, there may be other types of assessments (e.g., open questions, oral exams). Finally, our study was conducted using assessments in English, limiting the extent to which the findings apply to programming assessments in other languages. This poses a limitation as programming education is a global endeavour, and a significant proportion of programming courses and resources are available in languages other than English.

\textbf{Prompt Engineering.}
The prompts employed in this study were carefully crafted following the best practices. However, it is important to acknowledge that our research did not explore the effects of prompt engineering, i.e., further fine-tuning the initial prompts. Engaging in prompt engineering could potentially lead to even stronger performance of the models. This unexplored area could limit the implications of our findings which should rather be interpreted as lower bound of the performance.

\textbf{Information on GPT Models.}
It is not well-known what data have been used during the models' training. This is important because LLMs such as the ones evaluated in this study have capacity to memorize the data seen during their training. Hence, in case the assessments would have been seen during the training our experiments would not be able to show the models' capabilities to pass the assessments. They would rather be a testament to their memorization abilities. While we can be reasonably sure that the assessments employed in this study were not seen during the training as they are not part of any publicly available data set, this is an important limitation one has to be aware of when evaluating the OpenaAI's GPT models. Additional limitation is the lack of publicized technical details about GPT-4. The rapid development and evolution of GPT models, coupled with a lack of available technical details, makes it challenging for researchers to reproduce our study. Therefore, the results should be interpreted with this lack of full transparency and reproducibility in mind.

\section{Conclusions and Future Work}
We analyzed the capabilities of the GPT-4 model in passing typical assessments, such as MCQ tests and coding exercises, in introductory and intermediate programming courses. The analysis is the needed response to the recent release of GPT-4 evaluating the extent to which previous findings regarding GPT-3 and 3.5 models are still relevant. The study highlights that the risk of learners becoming overly reliant on GPT models when completing programming course assignments and assessments is a genuine concern that must be taken seriously which is consistent with \cite{becker2022programming}. In light of these findings, it is crucial to develop strategies to address this growing challenge and maintain the relevance and integrity of programming education.

The future work should focus on development of innovative assessment techniques resilient to automatically generated solutions. This could include incorporating real-time problem-solving components, group projects, or other collaborative activities that require human interaction. Additionally, further studies of potential benefits and risks associated with LLMs are needed to enable educators to harness their power while mitigating potential drawbacks. Finally, as more capable LLMs continue to emerge, it is crucial to conduct ongoing evaluations of their capabilities in the context of programming education. This will ensure that educators and institutions remain informed and prepared to adapt their teaching methodologies and assessment strategies in response to the rapid advancements.


\bibliographystyle{ACM-Reference-Format}
\balance
\bibliography{sample-base}

\newpage

\appendix

\section{MCQ Examples per Types}
\label{app:mcqs}

\subsection*{True/False}
\noindent Developers that write code individually are not expected to apply code standards.\\
A. True\\
\emph{B. False}\\

\noindent Evaluate the following expression and determine whether it is True or False.
\begin{verbatim}2 + 2 != 2 * 2\end{verbatim}
A. True\\
\emph{B. False}

\subsection*{Identify True/False Statement}
Which of the following statements is false?\\
A. The pandas module provides some CSV-related methods.\\
B. Python has a built-in XML package with several modules for XML parsing.\\
\emph{C. JSON data format has syntax to represent all Python data structure types.}\\
D. Python has a built-in \verb|csv| module containing methods for reading and writing into CSV files.\\

\noindent Take a look at the snippet and choose one of the following statements which is true:
\begin{verbatim}
nums = []
vals = nums[:]
vals.append(1)
\end{verbatim}
A. \verb|nums| is longer than `vals`\\
\emph{B.} \verb|vals| \emph{is longer than} \verb|nums|\\
C. \verb|nums| and \verb|vals| are of the same length

\subsection*{Finish Statement}
The `**` operator:\\
A. performs duplicated multiplication\\
B. does not exist\\
\emph{C. performs exponentiation}\\

\noindent Right-sided binding means that the following expression:
\begin{verbatim}1 ** 2 ** 3\end{verbatim}
will be evaluated:\\
\emph{A. from right to left}\\
B. in random order\\
C. from left to right

\subsection*{Output}
What is the output of the following snippet if the user enters two lines containing \verb|2| \emph{and} \verb|4| \emph{respectively?}
\begin{verbatim}
x = int(input())
y = int(input())
print(x + y)\end{verbatim}
A. 2\\
B. 24\\
\emph{C. 6}\\

\noindent What is the output of the following snippet?
\begin{verbatim}
my_list_1 = [1, 2, 3]
my_list_2 = []
for v in my_list_1:
    my_list_2.insert(0, v)
print(my_list_2)\end{verbatim}
A. \verb|[1, 2, 3]|\\
B. \verb|[1, 1, 1]|\\
C. \verb|[3, 3, 3]|\\
\emph{D.} \verb|[3, 2, 1]|

\subsection*{Fill-in Blanks}
Fill in the blank of the \verb|is_negative| function definition shown below, so that the function returns \verb|True| when the argument provided to \verb|num| is a negative number and returns \verb|False| otherwise.
\begin{verbatim}
def is_negative(num):
    return _________________
\end{verbatim}
A. \verb|not (num > 0)|\\
B. \verb|num > 0|\\
C. \verb|num <= 0|\\
\emph{D.} \verb|num < 0|\\

\noindent The following code snippet should open the \verb|myfile| file and assign the lines to the \verb|all_lines| variable. Which of the options below should be used to fill in the blanks?
\begin{verbatim}
with __________________________
    all_lines = file.readlines()\end{verbatim}
\emph{A.} {\ttfamily \emph{open("myfile",'r') as file:}}\\
B. \verb|"myfile" in open as file:|\\
C. \verb|with open "myfile" as file:|

\subsection*{Other}
How many times will the code snippet below print `X`?
 
\begin{verbatim}
for i in range(1, 7):
    for j in range(2, 6):
    print('X')\end{verbatim}
\emph{A. 24}\\
B. 28\\
C. 35\\

\noindent Given the piece of code presented in the code snippet below, what is the value of \verb|palindromes[1]|?
\begin{verbatim}
palindromes = ['pop', 'noon', 'madam']
\end{verbatim}
A. \verb|'pop'|\\
\emph{B.} \verb|'noon'|\\
C. \verb|'p'|\\
D. \verb|'madam'|\\
E. \verb|'o'|








\end{document}